\documentclass{aa}
\usepackage{graphicx}
\usepackage[varg]{txfonts}

\newcommand{\Mpcinvh}{\ensuremath{h{\rm Mpc^{-1}}}}

\usepackage{natbib,twoopt}
\usepackage[breaklinks=true]{hyperref} 
\bibpunct{(}{)}{;}{a}{}{,}          
\makeatletter
  \newcommandtwoopt{\citeads}[3][][]{\href{http://adsabs.harvard.edu/abs/#3}%
    {\def\hyper@linkstart##1##2{}%
     \let\hyper@linkend\@empty\citealp[#1][#2]{#3}}}
  \newcommandtwoopt{\citepads}[3][][]{\href{http://adsabs.harvard.edu/abs/#3}%
    {\def\hyper@linkstart##1##2{}%
     \let\hyper@linkend\@empty\citep[#1][#2]{#3}}}
  \newcommandtwoopt{\citetads}[3][][]{\href{http://adsabs.harvard.edu/abs/#3}%
    {\def\hyper@linkstart##1##2{}%
     \let\hyper@linkend\@empty\citet[#1][#2]{#3}}}
  \newcommandtwoopt{\citeyearads}[3][][]%
    {\href{http://adsabs.harvard.edu/abs/#3}
    {\def\hyper@linkstart##1##2{}%
     \let\hyper@linkend\@empty\citeyear[#1][#2]{#3}}}
\makeatother

\usepackage{xspace}
\newcommand{\hs}{{\sc haloscope}\xspace}

\bibpunct{(}{)}{;}{a}{}{,} 

\begin{document}

\title{The multi-dimensional halo assembly bias can be preserved when enhancing halo properties with \hs}

\author{Sujatha~Ramakrishnan \inst{\ref{inst1},\ref{inst2},\ref{inst3}}
\and Violeta~Gonzalez-Perez \inst{\ref{inst1},\ref{inst2}}
\and Gabriele Parimbelli \inst{\ref{inst3}}
\and Gustavo~Yepes\inst{\ref{inst1},\ref{inst2}}}
\institute{Departamento de F\'isica Te\'orica,  Facultad de Ciencias M-8,   Universidad Aut\'onoma de Madrid, 28049 Madrid, Spain \label{inst1}
\and Centro de Investigaci\'{o}n Avanzada en F\'isica Fundamental (CIAFF), Facultad de Ciencias, Universidad Aut\'{o}noma de Madrid, 28049 Madrid, Spain\label{inst2}
\and Institute of Space Sciences (ICE, CSIC), Campus UAB, Carrer de Can Magrans, s/n, 08193 Barcelona, Spain \label{inst3}}

\titlerunning{Preserving multi-dimensional assembly bias with \hs}
\authorrunning{Ramakrishnan et al. }
\date{15 November 2024 /
14 March 2025 }
\abstract {Over $90$\% of dark matter haloes in cosmological simulations have unresolved properties. This can hinder the dynamical range of simulations and result in systematic biases when modelling cosmological tracers. Current methods for enhancing unresolved haloes cannot preserve the multi-dimensional assembly bias found in simulations.}
{We aim to more precisely determine unresolved structural and dynamical halo properties while preserving the correlations with environment and halo assembly bias found in simulations.}
{We have developed \hs, a machine learning technique that uses multi-variate conditional probability distribution functions. This method ensures that correlations among various halo properties, as well as their dependence on the local environment, are preserved. In this work, we trained \hs with a high-resolution (HR) simulation and used it to better determine the properties (concentration, spin, and two shape parameters) of unresolved dark matter haloes in an eight times lower resolution simulation.}
{\hs is able to recover the multi-dimensional halo assembly bias, that is, the correlations of different combinations of halo properties with the large-scale environment, measured in the HR simulation. This is achieved by including the linear halo-by-halo bias and tidal anisotropy in the set of input training parameters. \hs, by design, also recovers the joint distribution of the halo properties.
To study how resolution effects propagate into the clustering of model galaxies, we generated catalogues of central galaxies using two implementations of the assembly bias in a halo occupation distribution model. The clustering of central model galaxies is improved by a factor of three at $0.009<k (\Mpcinvh)<0.6$ when the unresolved haloes are enhanced with \hs.}
{Our method can preserve the multi-dimensional halo assembly bias when trained using the local environment of haloes. \hs can improve the accuracy of cosmological tracer catalogues produced with approximate methods when many realisations are needed.
}

\maketitle

\section{Introduction}

Over $80$\% of dark matter haloes in numerical simulations have unresolved halo masses, and this number increases to $90$\% when considering other halo properties (see Fig. \ref{fig:fraclowres}). Simulations are required to assess the systematic errors and incompleteness of current and future cosmological surveys \citetext{\citealt{2009arXiv0912.0201L,2024arXiv240513491E} and \citealt{desiSV_2024}}. These surveys are probing the large-scale structure of the Universe to understand the nature of dark matter and dark energy. The tailored simulations that support these surveys \citetext{e.g. \textit{Euclid} Flagship, Outer Rim, and Abacus Summit; \citealt{2017ComAC...4....2P}, \citealt{2019ApJS..245...16H}, and \citealt{2021MNRAS.508.4017M}}
 are hindered by their unresolved halo properties \citep{2021MNRAS.500.3309M}.

The most common way to overcome unresolved halo properties is to resample the properties of low-mass haloes with fitting functions that provide the mean and scatter of a given halo property \citetext{\citealt{2008ApJ...678..621K},\citealt{2015ApJ...799..108D}, \citealt{2019ApJ...871..168D} and \citealt{Ishiyama:2020vao}}. However, correlations between halo properties have only been taken into account in a handful of studies \citetext{e.g. \citealt{farahi2022} and \citealt{mendoza2023}}. None of the current methods for more precisely and accurately determining unresolved halo properties can preserve the multi-dimensional assembly bias (AB) found in simulations. 

Halo AB is a term used to describe the residual dependence of the large-scale clustering on halo properties other than mass after fixing halo mass.\citetext{\citealt{2004MNRAS.350.1385S},\citealt{2005MNRAS.363L..66G},\citealt{2010ApJ...708..469F} and reference therein}. Many studies over the past decades have helped develop the understanding that the large-scale halo AB is a consequence of the tidal influence on the halo properties at relatively small scales \citetext{\citealt{2007MNRAS.375..489H},\citealt{2015ApJ...807...37S},\citealt{2017MNRAS.469..594B},\citealt{Salcedo:2017ndf},\citealt{2018MNRAS.476.4877M},\citealt{2019MNRAS.489.2977R}}. 

The halo AB affects the clustering of the galaxies they host. Galaxy AB studies based on both galaxy models and hydrodynamical simulations have determined that the local environment influences galaxy clustering \citep{vgp2020,2021MNRAS.502.3242X,2024A&A...685A..61B,2024A&A...690A.221P,2024MNRAS.527.3771A,2024MNRAS.530..947Y}.
The halo environment at large and intermediary scales has been used to generate galaxy catalogues that can recover the two-point correlation function measured by SDSS-IV/eBOSS, GAMA, and DESI \citetext{\citealt{2024A&A...690A.221P},\citealt{2024MNRAS.527.3771A} and \citealt{2024MNRAS.530..947Y}} or the bispectrum \citep{2024JCAP...07..083C}. 

Halo occupation distribution (HOD) models have been widely used to explore the effect that uncertainties in the clustering of particular cosmological tracers might have on cosmological inferences \citep[e.g.][]{Alam_2021}. The simplest HOD models place galaxies into haloes based only on their mass \citetext{\citealt{Zheng2005} and \citealt{2011ApJ...736...59Z}}. However, several studies highlight the requirement to go beyond halo mass to reproduce the observed clustering of cosmological tracers, such as galaxies \citetext{\citealt{2015MNRAS.446..651W},\citealt{2016MNRAS.460.2552H},\citealt{2018MNRAS.478.4487T} and \citealt{rocher23}}. 

\citet{2021MNRAS.500.1029L} and \citet{2024MNRAS.530.3127Z} point out the need to take into account the various correlations between halo properties, in  addition to the environment, when modelling clustering systematic errors. Besides occupation, models for the galaxy-halo connection have also explored associating galaxy properties such as galaxy spins, bar formation, and disk size with halo properties \citetext{\citealt{1980MNRAS.193..189F},\citealt{1998MNRAS.295..319M} and \citealt{2022ApJ...940..175K}}

In this work we used the halo mass and local environment to develop a multi-dimensional prediction for the internal halo properties (concentration, shape, and spin). Previously, we made this prediction for one halo property at a time at a given redshift \citep{2021MNRAS.503.2053R} and incorporating a cosmology and redshift dependence \citep{ramakrishnan2022}. Here, we extend the capability of our algorithms to incorporate correlations between the halo properties and to be able to recover the multi-dimensional halo AB (i.e. the dependence of the halo bias and clustering on two or more halo properties). We also applied them to a model galaxy population built on a low-resolution (LR) halo catalogue and show the effects and improvements in the predicted galaxy clustering.

The paper is organised as follows. Section~\ref{sec:sim} describes the specifications of the set of simulations that are used in this paper. It also introduces the dark matter haloes in the simulations and their properties. Section~\ref{sec:method} introduces the mathematical framework of our algorithm, \hs\ (HALO propertieS having COvariance Preserved with Environment), used to predict the multi-dimensional distribution and AB of halo properties. This code is publicly available\footnote{\url{https://github.com/computationalAstroUAM/haloscope}}. In Sect.~\ref{sec:resultshalo} we show how \hs can be used on poorly resolved haloes from a LR simulation to recover the properties and associated correlations of haloes from a high-resolution (HR) simulation. In Sect.~\ref{sec:galaxy}, using an HOD with AB modelling, we show how poorly resolved haloes can affect the modelling of galaxy clustering and how our algorithm can be used to improve it. We conclude in Sect.~\ref{sec:conclusion}.

\section{Simulations}
\label{sec:sim}
Our primary N-body simulation suite is the {\sc unit}\footnote{\url{http://www.unitsims.org/}} \citep{Chuang2019}, which has a box size of $1 {\rm Gpc} h^{-1}$. These are dark-matter only runs with the following cosmological parameters: $\Omega_{m} = 0.3089,~ h_{0}=0.6774,~n_{s}=0.9667,~\sigma_{8}=0.8147$. We used a pair of simulations with two dark-matter particle mass resolutions, one with $m_p= 1.2 \times 10^{9} M_{\odot}h^{-1}$, which we call {a HR simulation}, and another with eight times worse resolution (i.e. $m_p= 9.6 \times 10^{9} M_{\odot}h^{-1}$), which we call {a LR simulation} . We performed our analysis on the final simulation snapshot, at $z=0$.
    
The dark matter haloes in the simulation were identified using \textsc{rockstar} \citep{2013ApJ...762..109B} and \textsc{consistent trees} \citep{2013ApJ...763...18B}. 
\subsection{Halo properties}
In our analysis we used several of the dark matter halo properties computed by \textsc{rockstar}. We divided this properties into primary and secondary properties:
\begin{itemize}
    \item Primary halo property: The primary property of dark matter haloes is their mass. By default, we used $M_{200b}$. This is the mass enclosed inside $200$ times the background density, $R_{200b}$. We discuss other possible primary halo properties in Appendix \ref{sec:alternatemass}. 
    
    \item Secondary halo properties:
    \begin{enumerate}
        \item $c_{vir}$ - is the slope of the Navarro-Frenk-White (NFW) density profile. It is also a proxy for merger history of the halo \citep{Wang:2020hpl}.
        \item $\lambda$ - measure of the angular momentum of the halo.
        \item $c/a$ - ratio of the smallest ellipsoidal axis to the largest.
        \item $b/a$ - this is the ratio of the second smallest ellipsoidal axis to the largest.
    \end{enumerate}
\end{itemize}

Before the analysis, we also applied cleaning cuts on the haloes to ensure that only parent haloes are considered $(\textsc{pid}=-1)$ and to select for virialised haloes ($2T/|U|\leq 2$) \citep{2007MNRAS.376..215B}. We also ensured that the haloes had not been subjected to recent major mergers, which could drastically alter their halo properties ($z_{\rm lmm}>0.4$). 

\subsection{Environmental properties}\label{sec:env}
The key idea of our method is to assign internal halo properties using descriptors of the halo's environment. These environmental descriptors need to be computed at sufficiently large scales that they will also be resolved by either lower resolution simulations or fast approximate methods. In this regard we used as input three halo environmental properties: tidal anisotropy ($\alpha$), overdensity ($\delta$), and the linear bias ($b_1$). Below we define these properties and we describe the scales at which they are computed.

\subsubsection{Tidal anisotropy}\label{sec:tidal_anisotropy}
We also describe the environment of each halo outside of its boundary at different scales. For this, we primarily used the overdensity of the halo and the tidal anisotropy of the halo. Both quantities were constructed using the eigenvalues $\lambda_{1},\lambda_{2},\lambda_{3}$ of the tidal tensor field $\partial_{i}\partial_{j} \psi$ at several smoothing scales \citetext{\citealt{10.1093/mnras/282.2.436},\citealt{10.1093/mnras/232.2.339} and \citealt{2007MNRAS.375..489H}}:
\begin{equation}
\delta_S = \lambda_{1}+\lambda_{2}+\lambda_{3}
\label{eq:delta}
\end{equation}
\begin{equation}
\alpha_S = \dfrac{\sqrt{q^{2}}}{1+\delta_{S}}
\label{eq:alpha}
,\end{equation}
where $q^2={(1/2)[(\lambda_{1}-\lambda_{2})^2+(\lambda_{2}-\lambda_{3})^2+(\lambda_{3}-\lambda_{1})^2}]$, and $S$ is the choice of smoothing scale. We refer to \citet{2018MNRAS.476.3631P} for the exact procedure to compute it. In \cite{2019MNRAS.489.2977R} it was statistically established that the tidal anisotropy $\alpha_{R_{\rm 200b}}$ \footnote{The above smoothing scale is for a top hat smoothing filter. This is equivalent to $4\times R_{\rm 200b}/\sqrt{5}$ for a Gaussian filter, which is what we used in practice. See Appendix A2 of \citet{2018MNRAS.476.3631P} for a discussion.} is the primary indicator of the halo AB. We computed $\alpha_{S}$ at 20 different Gaussian smoothing scales ranging from $0.24$ to $4 {\rm Mpc} h^{-1}$ and interpolated in between to assign for each halo the $\alpha_{4R_{200b}}$.  The choice of the smallest and the largest smoothing scales are proportional to the smallest and largest haloes in the {\sc unit} simulation. In Appendix~\ref{app:justification_for_smoothing_scale} we provide an independent justification for this choice of smoothing scale as it maximises correlation with the overdensity when compared to all other smoothing scales (see Fig.~\ref{fig:justification_smoothingscale}).
A multi-scale treatment of the cosmic web filaments is essential for accurately describing the environment of the dark matter halo \citetext{\citealt{2021MNRAS.502.5210P} and \citealt{2024JCAP...09..041D}}.

\subsubsection{Overdensity}\label{sec:overdensity}
The overdensity, $\delta_{10 R_{200b}}$, has already been defined in terms of the eigenvalues of the tidal tensor in Eq. \ref{eq:delta} in the section above. This is computed at a larger smoothing scale of about $10 \times R_{200b}$. 

\subsubsection{Linear bias}\label{sec:b1}
 We computed the linear bias, $b_1$, at large scales, $r>60 h^{-1}{\rm Mpc}$ ($k_{max} = 0.1 h Mpc^{-1}$). For this calculation, we followed the halo-by-halo bias estimator introduced in \citet{2018MNRAS.476.3631P}\citetext{see also \citealt{2019MNRAS.489.2977R},\citealt{2021MNRAS.504.5205C},\citealt{2024A&A...692A..32B}}. 
This estimator provides a single value for the linear bias of each halo. This halo-by-halo bias $b_1$ is defined as follows,
\begin{equation}
b_{1} = \sum_{\rm low - k} w_{k}\left[\langle e^{i.\vec{k}.\vec{x}} \delta^{*}(\vec{k})\rangle_{k}/P_{\rm mm}(k) \right] \,
,\end{equation}
where $\vec{x}$ is the spatial location of the halo whose bias is being computed, $\delta (\vec{k})$ is the density contrast of the dark matter field, $P_{mm}(k)$ is the matter power spectrum and $w_{k}$ are weights proportional to the $k$ modes available in the simulation. 

For a halo population, the average value of the halo-by-halo bias is equal to the linear bias obtained using a traditional estimator. In this case, the traditional estimator would be the ratio between the halo-matter power spectrum and the matter-matter power spectrum at $k<0.1 h Mpc^{-1}$.

\begin{figure*}
\sidecaption
 \includegraphics[width=12cm]{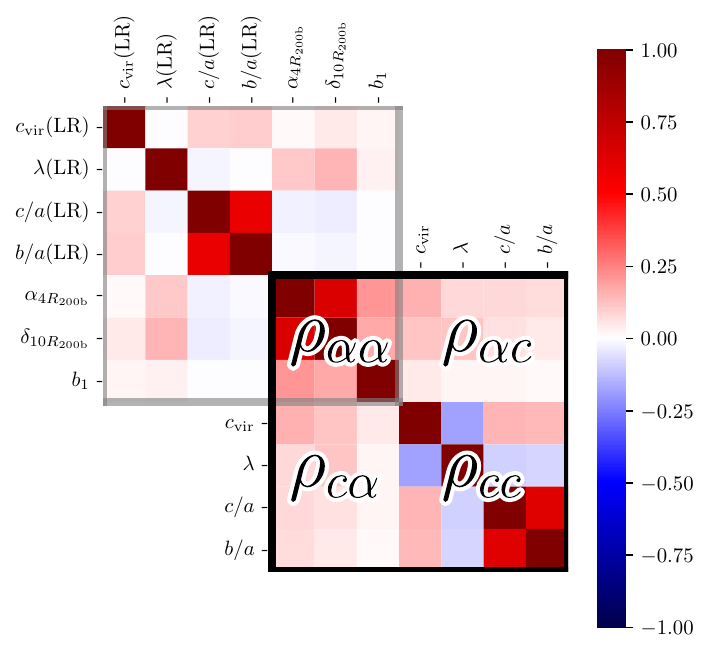}
\caption{Correlation matrices for the properties of haloes with the lowest masses considered here, $5\cdot 10^{11}<{\rm M_h} (M_{\odot}h^{-1})<7\cdot 10^{11}$. Each matrix element is colour coded with the value of the Spearman rank correlation value, from $-1$ to $1$, as indicated in the colour bar. We distinguish between intrinsic halo properties, $\vec{c}=\{c_{vir},\lambda,c/a,b/a\}$, and environmental ones, $\vec{\alpha} = \{b_{1},\alpha_{4R},\delta_{10R}\}$. We can understand these correlation matrices as being composed of four blocks (Eq. \ref{eq:blockmatrix}): correlations between environmental properties, $\tens{\rho_{\vec{\alpha} \vec{\alpha}}}$; correlations between intrinsic and environmental properties, $\tens{\rho}_{\vec{\alpha}\vec{c}}$ and $\tens{\rho}_{\vec{c}\vec{\alpha}}$; and correlations between intrinsic halo properties, $\tens{\rho}_{\vec{c}\vec{c}}$. The LR simulation matrix is at the top, outlined in grey, and the HR one is at the bottom, outlined in black. Since the environmental properties are well resolved in both simulations, we indicate a common correlation block matrix, $\tens{\rho_{\vec{\alpha}\vec{\alpha}}}$. Correlations between different halo properties are affected by resolution effects, as can be seen by comparing the top LR matrix with the bottom HR one. {The key idea of our method is to incorporate the missing correlations, i.e. the shades of blue, which indicate negative correlations present in the HR simulation (bottom matrix) but not in the LR simulation (top matrix).} 
}
\label{fig:corrmat}
\end{figure*}
\section{Methods: \hs}\label{sec:method}

Over $90$\% of haloes have unresolved properties (see Appendix~\ref{sec:motiv}) that can be improved. 
We developed \hs to improve unresolved halo properties. \hs is a machine learning (ML) technique that uses multi-variate Gaussian distributions with conditional probability given a set of halo properties. Our aims are to (i) impose halo property correlations and (ii) make an adequate choice of training parameters to preserve the multi-dimensional AB. 

We enhanced the secondary properties of LR haloes by training our algorithm with the properties of HR ones, whose mass resolution is eight times better (Sect. \ref{sec:sim}). Our method was applied and refitted in different mass bins.  

We consider the primary property of haloes to be their mass. Our algorithm does not modify this primary property and we used $M_{200b}$ (see Appendix \ref{sec:alternatemass} for a discussion on different halo mass definitions). There are several studies aiming at improving the halo mass of unresolved haloes \citep[e.g.][]{2022MNRAS.513.4318F}. The improvement achieved for halo masses using different techniques is marginal for the simulations under study (Appendix \ref{sec:improve_mass}). Thus, we proceeded to improving only secondary properties of LR haloes. 

Here we refer to the information we want to predict accurately as the { target vector}. This vector comprises various secondary structural and dynamical properties, $\vec{c}=\{c_{1},c_{2},..c_{r}\}$. The { input information} required by \hs is a vector containing two type of properties:
\begin{enumerate}[(i)]
\item Descriptors of the local density field around the halo $\vec{\alpha}=\{\alpha_{1},\alpha_{2},...\alpha_{q}\}$. The local density field is defined at a larger scale than the halo and hence sufficiently resolved within the LR simulation. 
\item The LR halo masses. Other properties from the LR are actually not important for the recovery of the HR secondary properties (this will be discussed in Sect. \ref{sec:featureimportance}).
\end{enumerate} 

We modelled the joint probability of the {target vector}, $\vec{c}$, and environmental properties, $\vec{\alpha}$, as a multi-dimensional Gaussian distribution, $\mathcal{N}$. This is a generalisation of the method described in \citet{2021MNRAS.503.2053R} to a multi-variate case.  If we have $r$ structural and dynamical halo properties $\vec{c}=\{c_{1},c_{2}...c_{r}\}$, and a vector of $q$ environment variables $\vec{\alpha}=\{\alpha_{1},\alpha_{2},...\alpha_{q}\}$ ; together they span a $(r+q)$ - dimensional Gaussian distribution:

\begin{equation}
\begin{bmatrix} \vec{c} \\ \vec{\alpha} \end{bmatrix} \sim \mathcal{N}\left( \begin{bmatrix}\vec{\mu}_{\vec{c}} \\ \vec{\mu}_{\vec{\alpha}}\end{bmatrix} , \begin{bmatrix}\tens{\sigma}_{\vec{c}} \tens{\rho}_{\vec{c}\vec{c}} \tens{\sigma}_{\vec{c}}& \tens{\sigma}_{\vec{c}} \tens{\rho}_{\vec{c}\vec{\alpha}} \tens{\sigma}_{\vec{\alpha}} \\  \tens{\sigma}_{\vec{\alpha}}\tens{\rho}_{\vec{\alpha}\vec{c}} \tens{\sigma}_{\vec{c}}  & 
\tens{\sigma}_{\vec{\alpha}} \tens{\rho}_{\vec{\alpha}\vec{\alpha}} \tens{\sigma}_{\vec{\alpha}}
\end{bmatrix}\right) \, .
\label{eq:blockmatrix}
\end{equation}

The multi-dimensional Gaussian above is characterised by a vector of means, $\vec{\mu}_{i}$, and standard deviations for each property in the diagonal elements of two block matrices,
$\tens{\sigma}_{\vec{c}} = {\rm diag}(\sigma_{c_{1}},\sigma_{c_{2}},...\sigma_{c_{r}})$ and  $\tens{\sigma}_{\vec{\alpha}} = {\rm diag}(\sigma_{\alpha_{1}},\sigma_{\alpha_{2}},...\sigma_{\alpha_{q}})$. The block matrix of correlation coefficients between the different internal halo properties is $\tens{\rho}_{\vec{c}\vec{c}} = [{\rho_{c_{i}c_{j}}}$]; the block matrix of correlation coefficients between the different environment variables is $\tens{\rho}_{\vec{\alpha}\vec{\alpha}} = [{\rho_{\alpha_{i}\alpha_{j}}}]$
; and the block matrices of cross-correlations between the halo properties and the environment are  $\tens{\rho}_{\vec{c}\vec{\alpha}} = [{\rho_{c_{i}\alpha_{j}}}]$ and $\tens{\rho}_{\vec{\alpha}\vec{c}} = [{\rho_{\alpha_{j}c_{i}}}]$. 

We visualise the block correlation matrices from Eq. \ref{eq:blockmatrix} in Fig. \ref{fig:corrmat}, where we use the N-body simulations under study. In this case, we took four halo properties into consideration, $\vec{c}=\{c_{vir},\lambda,c/a,b/a\}$, and three halo environmental properties, $\vec{\alpha} = \{b_{1},\alpha_{4R},\delta_{10R}\}$. We then computed the block diagonal elements shown in Eq.~\ref{eq:blockmatrix}  (i.e. $\tens{\rho}_{\vec{\alpha}\vec{\alpha}}$, $\tens{\rho}_{\vec{\alpha}\vec{c}}$, and $\tens{\rho}_{\vec{c}\vec{c}}$) for the well-resolved HR simulations (demarcated by a black border in Fig. \ref{fig:corrmat}). Each element of the block matrices in Fig. \ref{fig:corrmat} is colour coded according to the Spearman rank correlation value, going from $-1$ to $1$. Such correlations between several halo properties have been studied previously \citep[e.g.][]{2023MNRAS.521.5570S} and we obtain qualitatively consistent values.  We also show a similar matrix for the LR simulation (top-left and demarcated by a grey border in Fig. \ref{fig:corrmat}). The block matrix of correlation coefficients between different environment variables, $\tens{\rho_{\vec{\alpha} \vec{\alpha}}}$, is well resolved in both the LR and HR simulations and is the same in the two because the simulations have similar initial conditions. Hence, $\tens{\rho_{\vec{\alpha} \vec{\alpha}}}$ is shown just once in Fig. \ref{fig:corrmat}, as a common overlapping region for both the LR (grey border) and HR (black border) simulations. 

Several correlations that exist for HR haloes are missing for their LR counterparts (Fig. \ref{fig:corrmat}). Notably,
 \begin{itemize}
     \item Negative correlations (blue shades within the black border in Fig. \ref{fig:corrmat}) between halo spin and other properties seen in the HR $\tens{\rho_{\vec{c}\vec{c}}}$ matrix, are absent in the LR one.
     \item The smaller Spearman rank correlation values in the LR $\tens{\rho_{\vec{c}\vec{\alpha}}}$ (more white and less intense red within the grey border in Fig. \ref{fig:corrmat}) indicate a loss of correlations with the environment (AB), when the resolution is lower. 
 \end{itemize} 

\hs aims to preserve the correlations present in the training set, the HR simulation in this case. Our model prediction for the distribution of $\vec{c}$ given $\vec{\alpha}$, $\vec{c}|\vec{\alpha}$, is given by a p-dimensional Gaussian distribution \citep[Chapter 2.3]{2007JEI....16d9901B}:
\begin{align}
\vec{c}|\vec{\alpha} &\sim p(c|\tilde{\alpha}) = \mathcal{N}({\vec{\bar{\mu}}},\tens{\bar{\Sigma}}) \, , 
\label{eq:model}
\end{align}
where the average mean is defined as
\begin{align}
{\vec{\bar{\mu}}} &= \vec{\mu}_{\vec{c}} + \tens{\sigma}_{\vec{c}}\tens{\rho}_{\vec{c}\vec{\alpha}}\tens{\rho}^{-1}_{{\vec{\alpha}}{\vec{\alpha}}} \tens{\sigma}^{-1}_{\vec{\alpha}}(\vec{\alpha}-\vec{\mu_{\vec{\alpha}}}) \, ,
\label{eq:mubar}
\end{align}
and the average scatter is
\begin{align}
\vec{\bar{\Sigma}} &= \tens{\sigma_{\vec{c}}}(\tens{\rho_{\vec{c}\vec{c}}}-\tens{\rho_{\vec{c}\vec{\alpha}}}\tens{\rho}^{-1}_{\vec{\alpha}\vec{ \alpha}}\tens{\rho_{\vec{\alpha} \vec{c}}})\tens{\sigma}_{\vec{c}} \,.
\label{eq:sigmabar}
\end{align}

In the case where $c$ and $\alpha$ are scalars, the terms in Eqs.~\ref{eq:mubar} and \ref{eq:sigmabar} simplify to $\tens{\rho_{{\vec{\alpha}}{\vec{\alpha}}}}=1$ and $\tens{\rho_{\vec{c}\vec{c}}}=1$. In this case, we can redefine $\tens{\rho_{\vec{c}\vec{\alpha}}}=\tens{\rho_{\vec{\alpha}\vec{c}}}=\rho_{c}$, and use the standardised $\alpha$ (i.e. $\tilde{\alpha}=(\alpha-\mu_{\alpha})\sigma^{-1}_{\alpha}$), reducing the above expressions to%
\begin{align}
\bar{{\mu}} &= \mu_c + \sigma_c\rho_c \tilde{\alpha} \\ 
{\bar{\Sigma}} &= {\sigma^2_{c}}(1-{\rho^2_{c}})\\
p(c|\tilde{\alpha})&=\dfrac{e^{-(c- \mu_c - \sigma_c\rho_c \tilde{\alpha})^2/2\sigma^2_{c}(1-\rho^2_{c})}}{\sqrt{2 \pi \sigma^2_{c}(1-\rho^2_{c})}} \, .
\end{align}
The above equation is similar to Eq. 4 in \citet{ramakrishnan2022}. The framework is also similar to that presented in  \citet[MultiCAM-with scatter]{mendoza2023} and \citet[with KLLR]{farahi2022}. \citet{mendoza2023} introduced a generalisation of the {conditional abundance matching} technique \citep[e.g.][]{2013MNRAS.435.1313H} and used the same formalism for predicting the present-day halo properties given the accretion history of the halo traced back in time. Here we target large-volume simulations and aim to improve the LR halo properties and their AB given the information about the present-day local environment. This highlights the diversity of possible applications of \hs to other problems in astronomy and cosmology.

\subsection{Transforming variables into Gaussians}\label{sec:gaussianisation}
The above formalism relies on the assumption that both the feature and target variables are Gaussian when in reality they have skewed distributions.
Hence, all the variables need to be transformed to have a Gaussian distribution before applying \hs, and then inverse-transformed to their original distributions.

Different transformations can be tailored to the distribution of each specific halo or environment property \citep{ramakrishnan2022}. For example, a logarithmic transformation works fairly well to transform halo concentration and halo spin to a Gaussian variable. However, this approach requires an analytical description of the distribution of halo properties. To overcome this issue, here we used instead the { quantile transformer} from {\sc sci-kit-learn}. This method is based on a multi-variate probability integral transformation \citep{10.1214/aoms/1177729394}. This transformation has two advantages: (i) it can transform any arbitrary distribution to a Gaussian distribution, and (ii) it preserves rank correlations between the variables.

\begin{figure*}
\sidecaption
  \includegraphics[width=0.7\linewidth]{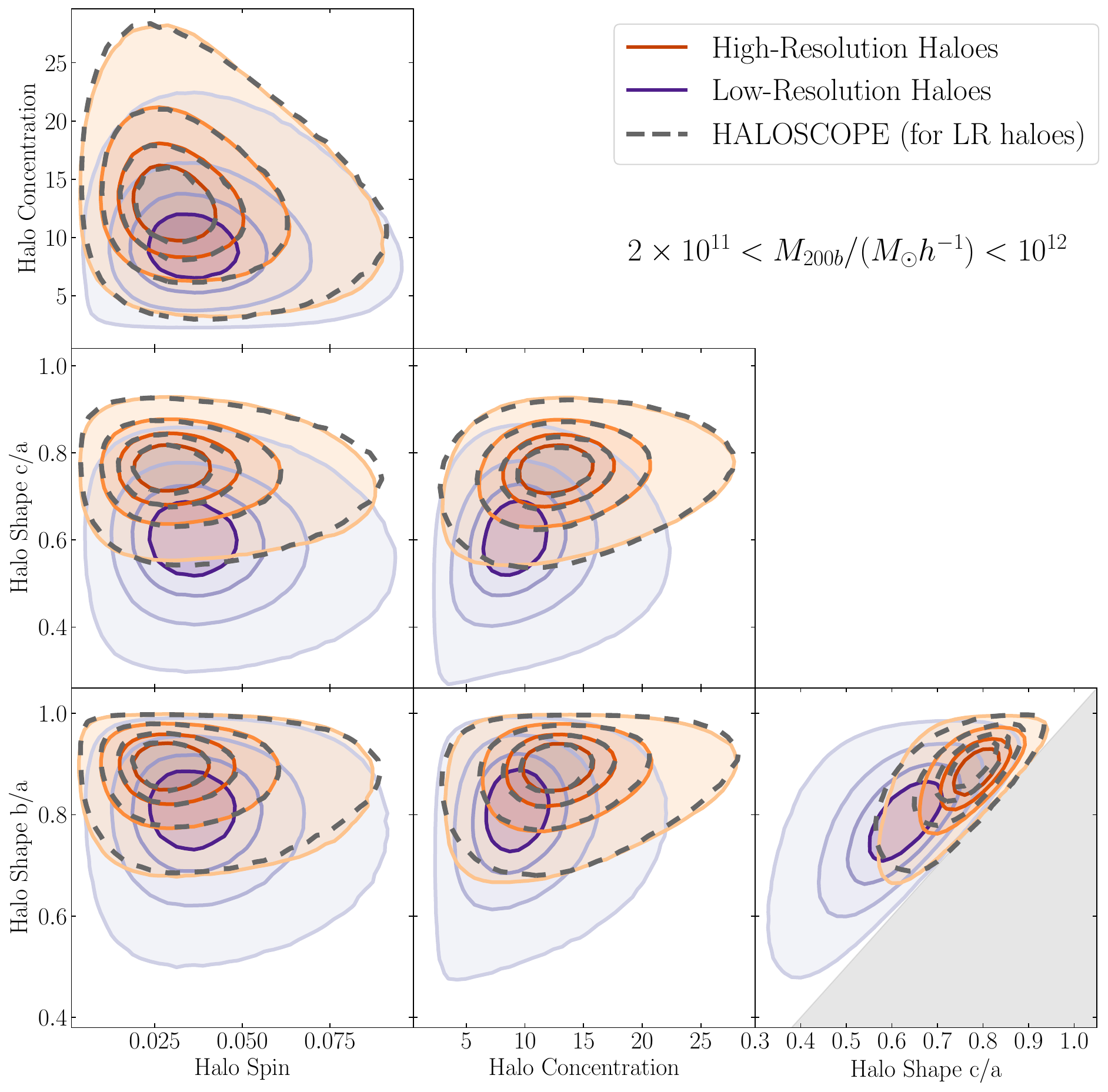}
     \caption{Confidence intervals (20\%,~40\%, 68\%, and 95\%) for the bi-variate distribution of pairs of halo properties in the mass range $2\times 10^{11}<M_{200b}(M_{\odot}h^{-1})<10^{12}$, corresponding to haloes having from $20$ to $100$ particles in the LR simulation. Purple contours show the distribution of LR haloes and orange the HR haloes. The dashed black contours represent the new distribution of the LR haloes after applying \hs. In the bottom right panel the grey area simply indicates a forbidden region due to the constraint $a>b>c$. 
      {Since our algorithm is designed to reproduce any multi-dimensional skewed distributions, it can also recover the bivariate distribution of LR halo properties as shown here.}
     }
     \label{fig:bivariate-distribution}
\end{figure*}
\subsection{Linear constraints}
Halo properties have the following physical constraints:
\begin{align}
c_{\rm vir} &> 0 , \\ \notag
\lambda &> 0,\\ \notag
c/a &> 0,\\ \notag
b/a - c/a &> 0 \\ \notag
1-b/a&>0. \notag 
\end{align}
These inequalities are not taken into account by the default sampling of a multi-variate Gaussian distribution described before. We solved this issue by using a rejection sampling algorithm that discards samples that fail to adhere to these constraints. We note that alternative methodologies for sampling a truncated Gaussian are available in the existing literature, owing to the ubiquity of such challenges in statistical applications.

For the case under study, rejection sampling has an acceptance rate greater than $96$\% for all the mass ranges considered. The shaded region in the bottom right panel in Fig. \ref{fig:bivariate-distribution} shows an example of the application of the rejection sampling algorithm.

\section{Recovering correlations between halo properties}\label{sec:resultshalo}

We used \hs to enhance the properties (concentration, spin and two shape parameters) of unresolved dark matter haloes in a LR simulation, given an eight times higher resolution (HR) one. By construction, the method is designed to recover the mean halo property relation with halo mass and its scatter (see Sect. \ref{sec:method} and Appendix \ref{sec:mean}). Besides this we are also interested in recovering the multi-dimensional distribution of halo properties at different mass ranges. This allows us to capture the self-correlations between halo properties that are present for HR but absent in the LR correlation matrix (see Fig. \ref{fig:corrmat}). Missing out on such correlations can impact the galaxy-halo connection models. For example, \citet{2020A&A...644A..76P} shows how residuals in galaxy scaling laws are sensitive to the anti-correlation between spin and concentration; and 
\citet{2024MNRAS.530.3127Z} models the correlations between halo properties to address systematics in galaxy cluster cosmology with weak lensing scaling relations.

We used our algorithm, \hs, to create a catalogue of enhanced halo properties for the LR simulation. Thus, the LR haloes have two sets of halo properties, one computed with the halo finder (LR) and the other using our algorithm (LR$+$\hs); we compared them with HR halo properties computed with a halo finder (HR). Our method is applied independently in halo mass bins. In particular, we used the mass bins indicated in \autoref{tab:kstest}, namely nine logarithmically spaced bins ranging from $2 \times 10^{11} h^{-1}M_{\odot}$ to $3.2 \times 10^{14} h^{-1}M_{\odot}$. These correspond to 20 particles to 30000 particles is the LR simulation and 160 to 240000 particles in the HR simulation. 

The bivariate distribution of a pair of halo properties, for haloes with $2\cdot 10^{11}<M_{200b}(M_{\odot}h^{-1})<10^{12}$, is shown using the 20\%, 40\%, 68\% and 95\% confidence contours\footnote{These confidence intervals roughly correspond to 0.25$\sigma$, 0.5$\sigma$, 1$\sigma,$ and 2$\sigma$ contours, respectively, for a 1D Gaussian distribution.} in Fig. \ref{fig:bivariate-distribution}. The HR distributions (orange contours Fig. \ref{fig:bivariate-distribution}) are clearly different from the LR ones (purple contours). The centre of the LR contours are below that for the HR halo properties, except for the spin. The same can be seen in the median relations at the lowest halo mass bin considered (see Fig. \ref{fig:prelimresult}). The shape of the LR contours are also different, more spherical and less tilted, than those for the HR haloes. As the halo mass increases, the LR distributions get closer to the HR ones (Fig. \ref{fig:bivariate-distribution-massbins}). This is expected, as the properties of massive enough haloes will be resolved also in the LR simulation.

The haloes from the LR simulations improved with \hs (grey dashed contours in Fig. \ref{fig:bivariate-distribution}) agree remarkably well with the HR halo properties. We quantified the goodness of fit for \hs using the Kolmogorov–Smirnov test for individual properties (Appendix \ref{sec:massbins}). We find a good agreement between the distributions of HR halo properties and the LR$+$\hs ones (\autoref{tab:kstest}).

The non-Gaussian shapes of the HR distributions can be recovered by \hs. This is possible thanks to the use of the { quantile transformer} from {\sc sci-kit-learn} in the final step from our method (Sect. \ref{sec:gaussianisation}). 

The median and scatter for each studied HR internal halo property is recovered by haloes improved by \hs (Fig. \ref{fig:prelimresult}) as a consequence of providing a good fit to the bi-variate correlations.

\begin{figure*}
    \centering
    \includegraphics[width=18cm]{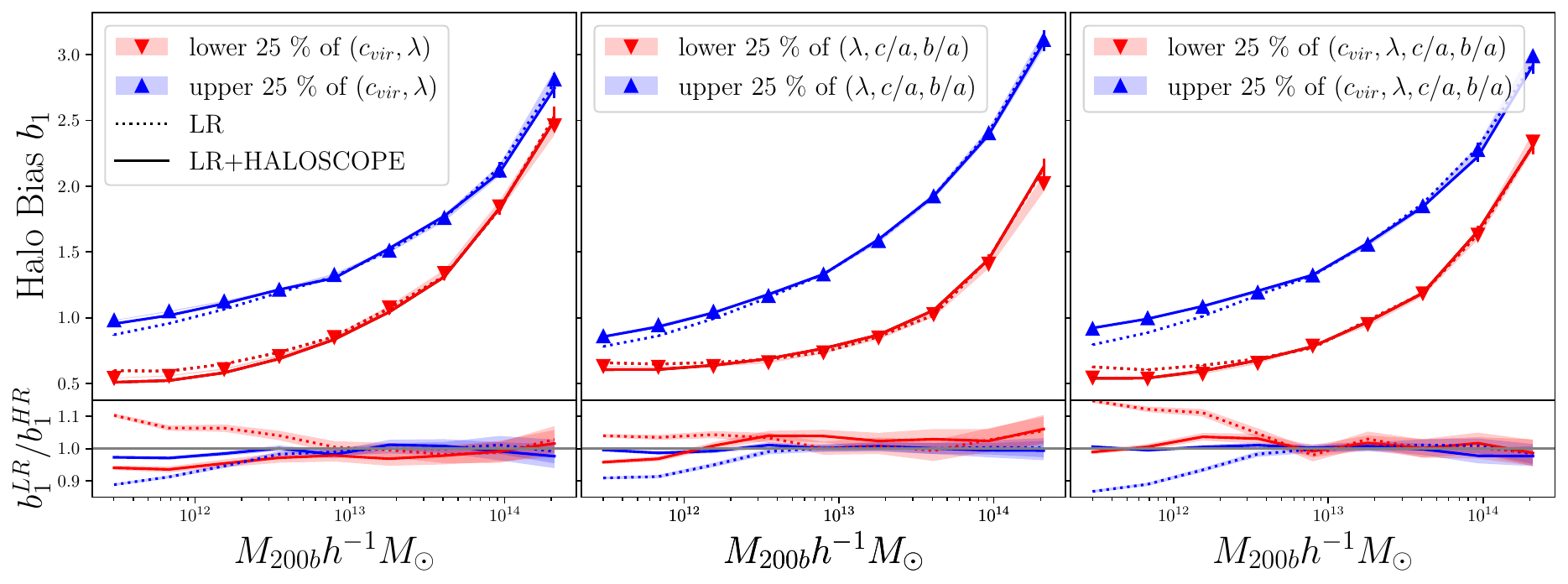}
    \caption{{ Top panels:} Linear halo bias as a function of halo mass for split populations of haloes, to show the halo AB. In blue (red) is shown the linear bias of the upper (lower) $25$\% of haloes according to: concentration and spin in the left panel; spin and two shape parameters in the middle panel; and concentration, spin, and two shape parameters in the right panel (see Sect.~\ref{sec:multiAB} for a detailed description of how the different halo populations are defined). Triangles show the HR haloes, dotted lines the LR ones, and the thick continuous lines are the result of applying \hs to the LR haloes, LR$+$\hs. Shaded regions correspond to the standard error of the mean. { Bottom panels:} LR and LR$+$\hs results divided by the HR ones. This shows that our method when applied to the LR haloes can decrease the differences between HR and LR from $\sim 12-15 \%$ to less than $5\%$.
    {Our algorithm is capable of recovering the multi-dimensional halo AB measured in HR simulations.}
    }
    \label{fig:ab}
\end{figure*}

\begin{figure*}
\includegraphics[width=18cm]{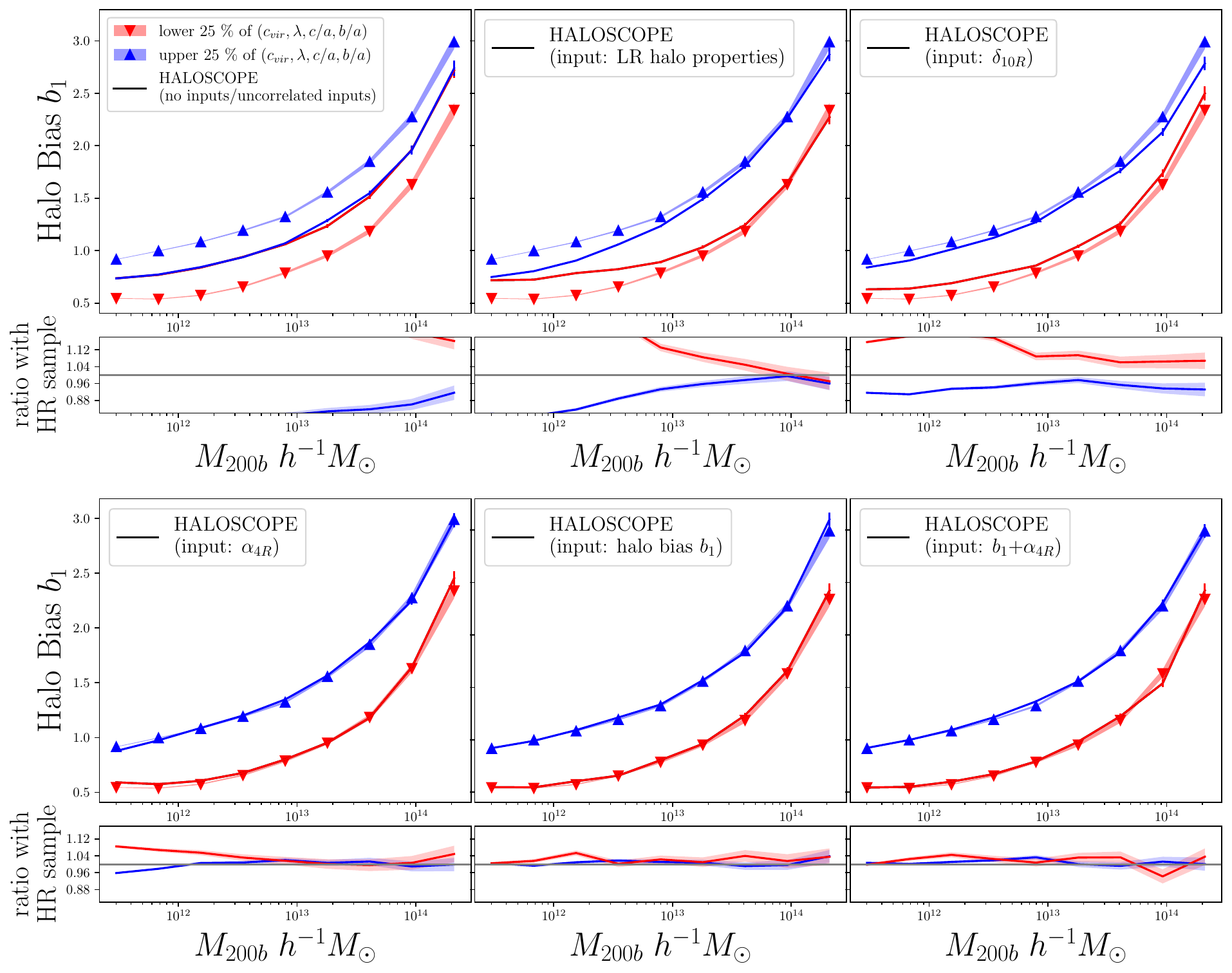}
\caption{Same as Fig. \ref{fig:ab}, but \hs has been 
 trained with different input properties. In the top left panel, random uncorrelated inputs are used for training \hs and when we apply our algorithm to the LR haloes, LR$+$\hs, no halo AB is measured. In the top middle panel, we train \hs with properties from LR haloes. For the other panels, \hs is trained with the input properties indicated in the legend. In all the panels, the blue and red triangles correspond to the upper and lower $25\%$ of ($c_{vir},\lambda,c/a,b/a$) HR haloes. This is our reference. 
{ To recover the multi-dimensional halo AB, \hs needs to be trained with haloes' environmental properties; in particular $b_1$, $\alpha_{4R}$, and a combination of the two give the best results.}
}
\label{fig:contributions}
\end{figure*}

\section{Recovering the multi-dimensional assembly bias}\label{sec:multiAB}
It is of paramount importance for any model catalogue to preserve the correlations between clustering statistics and the local density environment. The clustering of different tracers is one of the most widely used tools for inferring cosmological parameters from spectroscopic surveys. In the case of dark matter halo properties, such as dependence on the environment at fixed halo mass is called halo AB and has been widely studied in the literature \citetext{e.g.\citealt{2006ApJ...652...71W},\citealt{Croton:2006ys},\citealt{2008ApJ...687...12D},\citealt{2008MNRAS.388..638D},\citealt{2010ApJ...708..469F} and \citealt{2024arXiv240903004O}}. The strength and shape of the halo AB vary depending on the halo properties and mass range under consideration. Studies on halo AB have mostly tried understanding one halo property at a time \citep[e.g.][]{2010ApJ...708..469F}. In the literature, there are few studies of the halo AB as a function of a two halo properties simultaneously \citep[e.g.][]{2017JCAP...03..059L}.

Here we aim to model the {multi-dimensional AB}, that is to say, the linear bias of haloes with a given mass but classified into different sub-populations demarcated in the multi-dimensional distribution of halo properties. We can only achieve this with an adequate choice of training parameters. We discuss in Sect. \ref{sec:featureimportance} that environmental properties, in particular $\alpha_{4R}$ and $b_1$ (defined in Sect. \ref{sec:env}), are needed in the training of \hs to recover the AB measured for HR haloes. 

We quantified the halo AB by measuring  the linear bias (Sect. \ref{sec:b1}) for haloes split by their internal properties within equal mass bins (Fig. \ref{fig:ab}). To study the multi-dimensional nature of the halo AB, we rank-ordered haloes using two $(c_{vir},\lambda$; left panel in Fig. \ref{fig:ab}), three $(\lambda,c/a,b/a$; middle panel), and four $(c_{vir},\lambda,c/a,b/a$; right panel) properties and considering the upper and lower $25$\% percentiles in each case. For example, the lower $25$\% of $(c_{vir},\lambda)$ was chosen by selecting a population of haloes below the $p^{th}$  percentile in both $c_{vir}$ and $\lambda$. The upper $25$\% of $(c_{vir},\lambda)$ haloes correspond to a population above the $100-p^{th}$ percentile \footnote{Note that the value $p$ needed to encompass the $25$\% of the total population varies with the properties used to rank order the haloes. This value becomes larger as we take combinations of more halo properties.}. Other combinations of halo properties have also been tested; however, we focused on this set to show a range of AB trend with halo mass using a different number of halo properties. 

When we segregate haloes by their internal properties, there is a gap between the linear bias in the two populations. This is clearly seen when comparing the upper and lower $25$\% HR haloes (blue and red triangles) for the three panels in Fig. \ref{fig:ab}. In the case of using $(c_{vir},\lambda)$ to segregate haloes, the gap between the two populations steadily decreases with halo mass. We note that this gap does not invert as in the case of using halo concentration alone to segregate haloes of the same mass \citep[e.g.][]{ramakrishnan2022}. The trend for the gap, when using three parameters $(\lambda,c/a,b/a)$ to segregate haloes of similar mass, is opposite from before. In this case, the gap between the linear bias in the two populations increases with increasing halo mass. Lastly, when considering four parameters $(c_{vir},\lambda,c/a,b/a)$ to segregate haloes, the gap in the linear bias between the two populations at every halo mass is fairly constant.
 
The trend of the AB signal with halo mass can decrease, increase or remain fairly constant, depending on which properties are chosen to segregate haloes at a given mass (Fig. \ref{fig:ab}). Thus, we can conclude that the clustering signal can present different behaviours as a function of halo mass, depending on what secondary halo properties, beyond the halo mass, are considered.

\subsection{Improving the LR assembly bias}
We segregated haloes in the LR simulation in the same way to that described above for the HR simulation. The linear bias for LR haloes shows an AB signal with similar trends with halo mass as those described for the HR haloes (dotted lines in Fig. \ref{fig:ab}). However, the HR and LR AB signals are different (up to $12$\%) for low mass haloes.

When we improve LR haloes with \hs (solid lines in Fig. \ref{fig:ab}) we are able to recover an AB signal within $5$\% of that measured for HR haloes. The improvement is the largest for low-mass haloes, with fewer than 500 particles, which constitute over $90$\% of the total haloes (Fig. \ref{fig:fraclowres}). It is remarkable that haloes improved with \hs recover those HR trends in the halo AB, for any combination of halo properties.

The ability of \hs to recover the multi-dimensional AB is particularly useful for applications using fast approximate simulations or catalogues \citetext{\citealt{2016MNRAS.463.2273F} and \citealt{2019MNRAS.483L..58B}} that cannot provide halo AB but do contain sufficient information to obtain it with our method. In particular, these type of approximate methods have halo masses and tidal environments, which are the only inputs our model needs (see Sect. \ref{sec:featureimportance} for a study of feature importance of the inputs).

\subsection{Training set and assembly bias}\label{sec:featureimportance}

We studied what properties are needed to train \hs to recover the HR multi-dimensional AB. By construction, \hs will provide the bivariate distributions from the internal halo properties, $\vec{c}$, in the training set. However, this is not the case for the AB signal. Halo catalogues modelled simply with a mean or median concentration-mass relation with scatter, do not carry halo the AB information seen for the HR haloes (triangles in Figs.~\ref{fig:ab} and \ref{fig:contributions}).

If no environmental input, $\vec{\alpha}$, is given in the training, there is no AB. This can be seen in the top left panel of Fig. \ref{fig:contributions}, where we compare the HR AB signals with that of LR$+$\hs, when the algorithm is trained with either no environmental properties or uncorrelated inputs. In this case, the two segregated halo populations, using four parameters $(c_{vir},\lambda,c/a,b/a)$, show the same signal. 

Improved haloes do present a reduced AB signal when LR halo properties are used to train \hs (top middle panel in Fig. \ref{fig:contributions}). This signal differs by more than $20$\% from the target HR one at low masses, and is worse than the original LR signal shown in the right panel of Fig. \ref{fig:ab}. This results improves when the environmental property $\delta_{10R}$ (defined in Sect. \ref{sec:overdensity}) is used to train \hs. However, only using this parameter is not enough to recover an AB signal within $5$\% of the HR one.

Environmental properties are needed during the training of \hs to recover the HR AB. In particular, to achieve an agreement with the HR signal,  for the training below $10$\%  we do need to use the tidal anisotropy, $\alpha_{4R}$ (defined in Sect. \ref{sec:tidal_anisotropy}), or the linear bias,  $b_1$ (defined in Sect. \ref{sec:b1}), or better, a combination of the two. The tidal anisotropy smoothed at scales $4 R_{200b}$, $\alpha_{4R}$, has be shown to be the primary indicator for halo AB \citep[see Fig A5 in][]{2019MNRAS.489.2977R}. Nevertheless, we also analysed the density at a $10 R_{200b}$ scale, $\delta_{10R}$, to study its performance in this particular context. The density computed on these scales is expected to correlate with large-scale halo bias, $b_{1}$. However, we find that that using either $b_{1}$ or $\delta_{10R}$ leads to a different AB signal (~Fig. \ref{fig:contributions}). This difference might be introduced by using a fixed scale, $60 h^{-1}{\rm Mpc}$, when computing the halo-by-halo-bias, $b_{1}$, while the $10R_{200b}$ smoothing scale used for $\delta_{10R}$ varies among different haloes.

In this study, our default \hs algorithm is trained using the three environmental properties discussed here: $b_1$, $\alpha_{4R}$ and $\delta_{10R}$.

\section{Effect of halo properties on galaxy clustering}
\label{sec:galaxy}
In this section we show the effect that improving halo properties with \hs has on the clustering of central galaxies. We chose to focus solely on central galaxies to reduce the number of free parameters in the HOD model used to populate with galaxies the {\sc unit} simulation (described in Sect. \ref{sec:hod}). This choice allows us to get a cleaner view of the effect that \hs has on the galaxy clustering at large scales (described in Sect. \ref{sec:subhod}). We defer to future works a detail study of satellite galaxies.

\subsection{Model galaxy catalogues}\label{sec:hod}
\begin{figure}
    \centering
\includegraphics[width=0.99\linewidth]{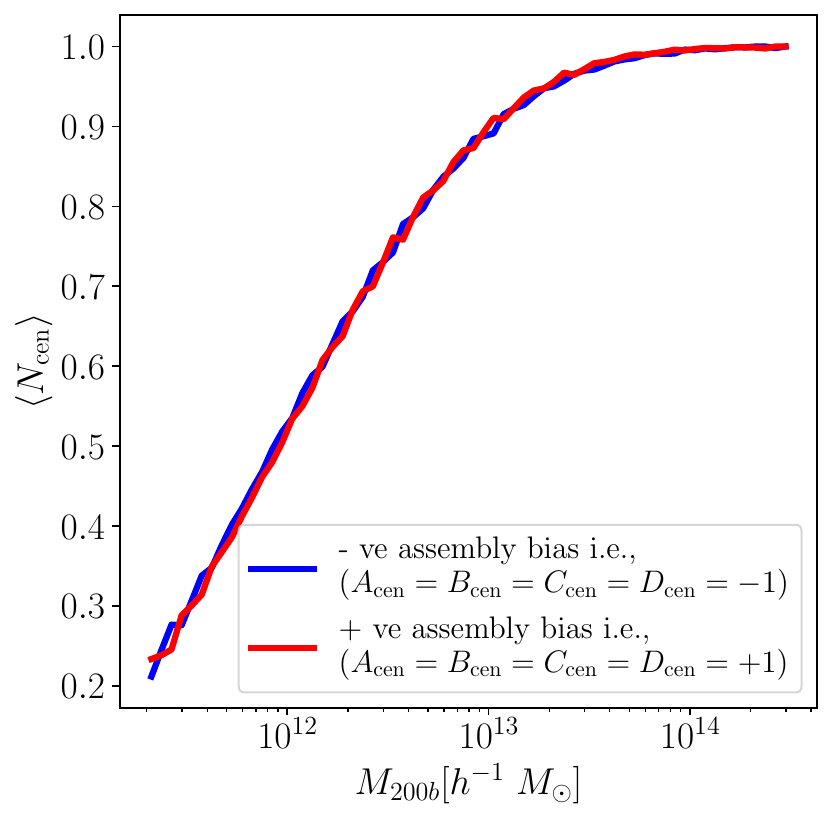}
    \caption{Average number of central galaxies as a function of halo mass for model galaxies generated using an HOD model with two different implementations of galaxy AB (Eqs.~\ref{eq:hod} and ~\ref{eq:gab}). 
    { We have produced two catalogues of central model galaxies hosted by haloes with similar mass distributions but different concentrations, spins, and shapes.}
    } 
    \label{fig:hod}
\end{figure}
\begin{figure}
    \centering
\includegraphics[width=0.99\linewidth]{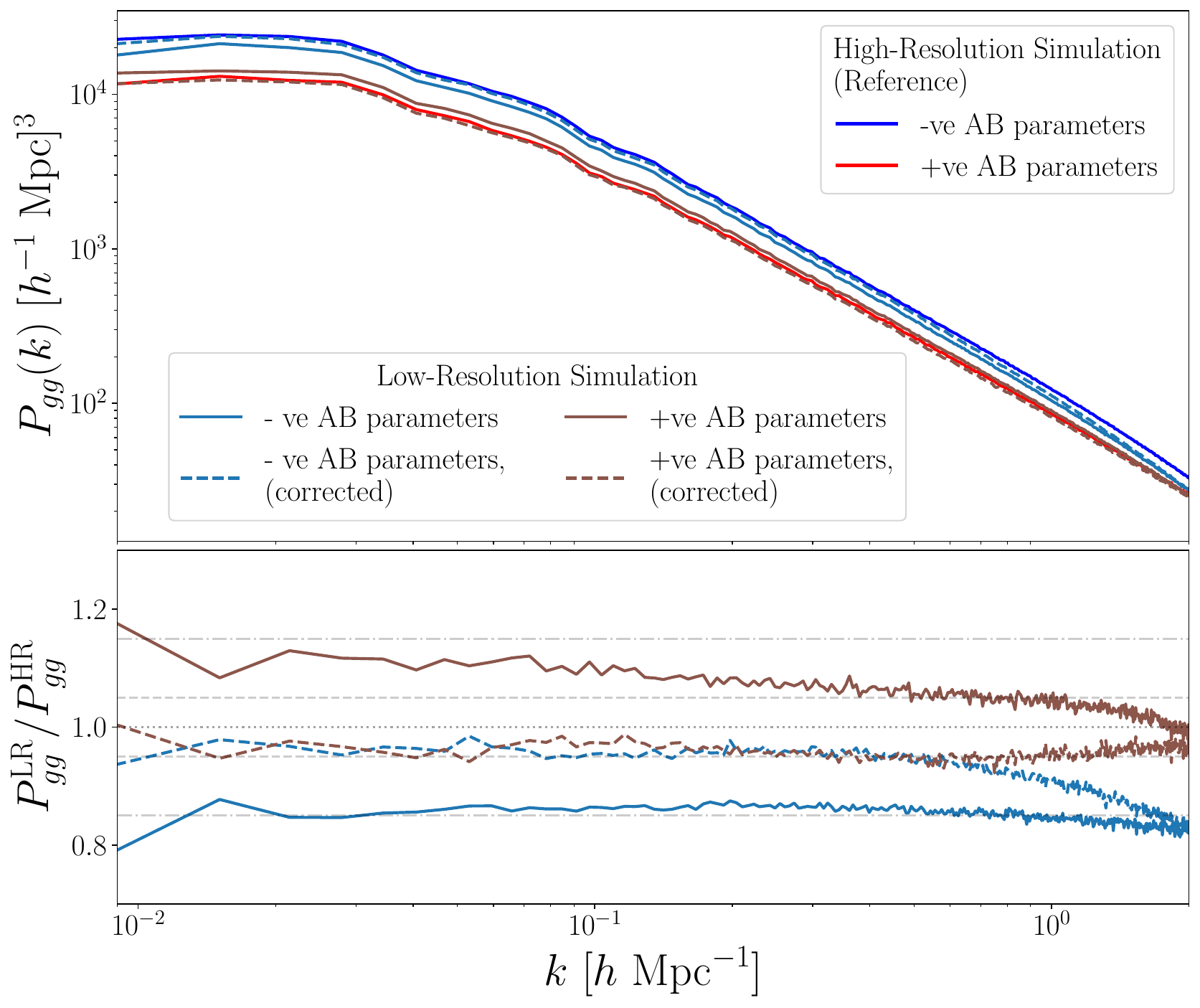}     \caption{{ Top panel:} Power spectrum for model central galaxies. In red are shown the power spectra measured in catalogues produced with a positive AB ({+ve AB}), with $A_{\rm cen} = B_{\rm cen} = C_{\rm cen} = D_{\rm cen} = 1$ in Eq. \ref{eq:gab}; in blue, those from catalogues with a negative AB ({-ve AB}), with $A_{\rm cen} = B_{\rm cen} = C_{\rm cen} = D_{\rm cen} = -1$. Continuous brighter lines show the results for galaxies generated on the HR simulation, while continuous lines with darker colours correspond to the LR simulation, as indicated in the legend. The dashed lines show the clustering from those catalogues generated on LR haloes corrected with \hs, LR$+$\hs.
{ There are differences between the power spectrum measured from galaxies within the {+ve AB} and {-ve AB} catalogues, due to galaxy AB.} These differences range from $20$ to $50$\% and are present in the whole range of $k$ modes.
{ Bottom panel:} Ratios with respect to the galaxy catalogues from the HR simulation of both the LR (solid lines) and LH$+$\hs (dashed lines) dark matter haloes. The $0$, $5,$ and $15$ percent levels are indicated by horizontal grey lines. The ratios of power spectra from two tracers in the same density field cancel the statistical noise in the bottom panel, the fixed-pair method used in the {\sc unit} simulations (more details in Sect.~\ref{sec:subhod}) also mitigates noise from cosmic variance. {There are $15$\% differences between the power spectrum of HR and LR galaxy catalogues. Our algorithm, can reduce this difference to $5$\%}.
}
\label{fig:galaxyps}
\end{figure}

We used an HOD model to populate with central galaxies the dark matter haloes of both the HR and LR simulations. We then studied the central galaxies clustering {for the same set of HOD parameters.} 
The continuous expansion in capabilities of cosmological surveys, such as DESI \citep{desiSV_2024} and \textit{Euclid} \citep{2024arXiv240513495E}, have created a demand for simulations that are larger in volume and more accurate at smaller scales. Faced with the computational costs of such simulations, to incorporate cosmological tracers in them, we often need to resort to connecting galaxies to haloes using computationally effective tools such as the HOD models. 

For the average occupancy of central galaxies in haloes of a given mass, $\langle N_{\rm cen} \rangle$, we used the following standard form \citetext{\citealt{Zheng2005} and \citealt{2024MNRAS.529.3877R}}:
\begin{equation}
\langle N_{\rm cen} \rangle = 1/2 \left( 1+ \mathrm{erf}\left(\dfrac{\log M_{h} - \log M_{\rm min}}{\sigma_{\log M}}\right)\right)  \, .
\label{eq:hod}
\end{equation}
We chose $\log M_{min} = 11.95$ and $\sigma_{\log M} = 0.65$ for our baseline model. This implies that $\langle N_{cen}\rangle$ transitions from $0$ to $1$ in a mass range centred in $\log M_{min} = 11.95$, as it can be seen in Fig. \ref{fig:hod}. For our LR simulation, this value corresponds to a few tens to hundreds of particles per halo. This makes our choice of parameters ideal for this specific study.

In addition, we also modelled the AB signal by altering $\log M_{min}$ \citetext{\citealt{2021MNRAS.502.3242X},\citealt{2023MNRAS.524.2507H} and \citealt{2024A&A...690A.221P}} in the following way,

\begin{equation}
\log M_{\rm min}^{\prime} = \log  M_{\rm min} + A_{\rm cen} f_{a} + B_{\rm cen} f_{b} + C_{\rm cen} f_{c} + D_{\rm cen} f_{d} \,
\label{eq:gab}
,\end{equation}
where $f_{a},f_{b},f_{c},f_{d}$ are, for a given mass range, the ranks a halo belongs to given the four properties under consideration ($a$,$b$,$c$ and $d$)\footnote{In practice it is preferable to model AB directly using halo environmental properties \citep[e.g.][]{2024MNRAS.527.3771A}, instead of internal ones. This reduces resolution effects. However, rank ordering by internal halo properties has been widely used in the literature \citep[e.g.][]{rocher23}}. The ranks are computed by first binning halo properties in narrow bins of halo masses, rank ordering them by increasing value of the halo property. The ranks are then rescaled to range between $-0.5$ and $0.5$. $A_{\rm cen}, B_{\rm cen}, C_{\rm cen}$, and $D_{\rm cen}$ are the AB parameters.

To explore the two extreme scenarios, we chose the following sets of AB parameters: the negative set $A_{\rm cen}=B_{\rm cen}=C_{\rm cen}=D_{\rm cen}=-1$; and the positive one ($A_{\rm cen}=B_{\rm cen}=C_{\rm cen}=D_{\rm cen}=1$. This choice also gives us a conservative estimate of the amount of systematic errors that can be introduced by resolution effects in a simulation.

We have created three pairs of galaxy catalogues, based on the HR, LR, or LR$+$\hs haloes. For each pair of catalogues, we have one with the negative set of AB parameters ({-ve AB}) and the other with the positive set of AB parameters ({+ve AB}). The average HOD is similar for the two model galaxy catalogues (Fig. \ref{fig:hod}), independently of the halo catalogue used. Catalogues produced with either a {-ve AB} or a {+ve AB} model, have similar distributions of central galaxies as a function of halo mass. 

The difference between the two sets of catalogues, with {-ve AB} or {+ve AB}, lie on the distribution of galaxies as a function of secondary halo properties. Central galaxies from the {-ve AB} catalogue are preferentially hosted by haloes that are less concentrated, have smaller spins and more spherical shapes, than those from the {+ve AB} catalogue. We expect these differences to propagate into the clustering of model galaxies.

\subsection{Clustering of model galaxies}
\label{sec:subhod}

We measured the power spectrum of the six galaxy catalogues we generated in pairs of {+ve AB} and {-ve AB}. The two sets of catalogues have similar average HODs (see Fig.~\ref{fig:hod}), independently from the underlying catalogue of dark matter haloes. We used the three halo catalogues available: directly the HR and LR simulations, and the LR haloes enhanced with other algorithm, LR$+$\hs.

There are differences of between $20$\% and $50$\% in the power spectrum measured from the pairs of {+ve AB} (solid red) and {-ve AB} catalogues (solid blue), as can be seen from the top panel in Fig. \ref{fig:galaxyps}. This result is independent of the input host haloes. The differences are found for all the $k$ modes considered. This shows that the halo AB directly propagates into the galaxy AB. Despite having similar average HODs (Fig. \ref{fig:hod}), model central galaxies show $\sim40$\% differences in their power spectrum for the pairs of {+ve AB} and {-ve AB} catalogues. Similar variations have been previously measured in studies of the galaxy AB signal as a function of multiple halo properties \citep{2024MNRAS.531..290M}.

The power spectrum from the LR galaxy catalogues differs by $15$\% from the HR version of the same HOD model (solid lines in the bottom panel of Fig. \ref{fig:galaxyps}). Such differences could possibly be absorbed into a different set of HOD model parameters. However, our analysis indicates that one has to be cautious when using the same HOD calibrations for simulations with a different mass resolution, in particular, when including the effect of AB. It should be noted that the differences between the LR and HR model catalogues have a reduced cosmic variance because of the fixed-pair technique used to generate initial conditions for the {\sc unit} simulations \citep{2016MNRAS.462L...1A}. Moreover, the LR and HR sets of the {\sc unit} simulations have identical initial phases, that is, they trace the same large-scale density field, so taking ratios is designed to cancel the statistical fluctuation errors \citep{2009JCAP...10..007M}. 

The power spectrum of the galaxy catalogues generated from LR haloes corrected by \hs, LR$+$\hs, is within  $5$\% of the HR one. Our method manages to reduce the previous discrepancy with LR by a factor of three (Fig. \ref{fig:galaxyps}). This agreement starts to deteriorate at small scales, $k>0.6 h Mpc^{-1}$, and the power spectra from LR$+$\hs catalogues agree with the LR ones by $k=2h Mpc^{-1}$. This seems to be a limitation of our method. 

Many HOD models address resolution issues by simply sampling from a fitting function such as a concentration-mass relation for poorly resolved haloes, and then rank ordering the samples to match the abundance of the existing poorly resolved sample \citetext{e.g. \citealt{Paranjape:2021zia} and \citealt{2024arXiv240513495E}}. 
Such an approach corrects for the value of the halo properties and incorporates AB to the levels expected in LR simulation. While such
a method corrects for the value of the halo properties and
incorporates AB to the levels expected in a LR simulation, we would like to emphasise that the
analysis here is more accurate since it incorporates AB to match the levels seen in well-resolved simulations.

The improvement in the clustering after applying \hs demonstrates the utility of our method to enhance summary statistics for model galaxies based on LR simulations.

\section{Summary and conclusions}
\label{sec:conclusion}
We have developed a ML method, \hs (Sect. \ref{sec:method}), to improve unresolved properties of dark matter haloes in simulations while preserving the multi-dimensional AB. The code associated with this method is publicly available\footnote{\url{https://github.com/computationalAstroUAM/haloscope}}. 

We have demonstrated the capabilities of \hs by accessing low-mass, poorly resolved haloes in a large-volume simulation and correcting for their halo properties; to produce results similar to those haloes that have achieved numerical convergence. In particular, we tested \hs using the original {\sc unit} simulation \citep{Chuang2019}, referred to here as the HR simulation, and a simulation with eight times worse resolution, the LR simulation (Sect. \ref{sec:sim}). For the application presented here, we aimed to recover the properties of the LR haloes. Our method was trained using halo properties in bins of halo mass. 

After training \hs with HR halo properties and then applying it to LR haloes, LR$+$\hs, we can conclude the following:
\begin{itemize}
\item \hs recovers the multi-variate distributions and correlations of halo properties in halo mass bins (Figs. \ref{fig:bivariate-distribution} and \ref{fig:bivariate-distribution-massbins}). It also recovers the median and scatter expected for the distribution of halo properties as a function of halo mass (Fig. \ref{fig:prelimresult}). 

\item \hs can recover, for the first time, the multi-dimensional halo AB (Fig. \ref{fig:ab}), that is to say, the simultaneous dependence of halo bias on arbitrary combinations of the halo properties at a fixed mass. This is only possible when environmental properties, such as the linear bias and the tidal anisotropy (Fig. \ref{fig:contributions}), are used to train \hs. LR$+$\hs reduces the differences in multi-dimensional AB between HR and LR from $12$\% to $5$\%.
\end{itemize}

This new method is particularly useful in the lowest mass range, where there are significant differences between LR and HR simulations. This is due to the large amount of unresolved properties for low-mass haloes in LR simulations. We have verified that our method and the conclusions above are robust against different definitions of halo mass (Appendix \ref{sec:alternatemass}). Our algorithm can also be applied after improving the halo mass function using different ML algorithms, such as random forest. However, this improvement provides very small changes to the halo mass that do not improve the unresolved properties of haloes, at least for the case under study (Appendix \ref{sec:improve_mass}).

The clustering of model galaxies can be affected by that of their host haloes. We studied this by generating catalogues of galaxies with an HOD model that includes AB. We generated catalogues of central galaxies based on the HR, LR, and LR$+$\hs haloes. We focused on central galaxies to be able to limit the free parameters of the HOD model. We used two extreme HOD models, one with a positive AB, {+ve AB}, and another with a negative one, {-ve AB} (see Eq. \ref{eq:gab}). After studying the power spectrum of the six galaxy catalogues, we conclude the following:
\begin{itemize}
    \item The galaxy AB is directly affected by the halo AB. The power spectrum differs by $20$-50\% (depending on the k mode) between the {\it +ve AB} and {-ve AB} galaxy catalogues  (Fig. \ref{fig:galaxyps}). This difference in clustering occurs despite the similar average occupancies with halo mass (Fig. \ref{fig:hod}).
    \item \hs improves the central galaxy power spectrum with respect to the HR catalogue (for $k<0.6 h^{-1} Mpc $): from a $15$\% difference with the LR catalogue to a $5$\% difference for the LR$+$\hs (Fig. \ref{fig:galaxyps}).
\end{itemize}

We expect our results to hold for satellites hosted by well-resolved haloes. The average HOD determines the number of satellites per halo. The number of satellite galaxies is typically expected to grow with the mass of their host halo. For tracers such as luminous red galaxies, satellite galaxies are expected to be hosted by haloes with masses above $10^{13}h^{-1}M_{\odot}$ \citep[e.g.][]{2024MNRAS.530..947Y}. These masses are well resolved by most current cosmological simulations and thus would not be affected by resolution effects at lower masses. In contrast, other tracers, such as emission line galaxies, are hosted by less massive dark matter haloes and have typically one or two satellite galaxies per halo \citetext{\citealt{vgp2020},\citealt{2024MNRAS.529.3877R} and \citealt{yu2024}}. From a modelling perspective, satellites residing in unresolved parent haloes can be influenced by resolution effects. In particular, unresolved internal properties of host haloes will impact the modelling of the radial distribution, velocities, and AB for satellites galaxies. Future studies should incorporate satellite galaxies to further test these assumptions.

\hs can serve as a unified framework for simulations with varying initial conditions and cosmologies if used in bins of peak height, instead of halo mass. This is because the correlation between internal properties of dark matter haloes and their environment is universal for variations around the standard cosmological model \citep{ramakrishnan2022}. One important aspect that remains to be tested is whether the correlations among the internal halo properties themselves are also universal across cosmologies and redshift. A confirmation of this universality would allow us to establish \hs as a fully cosmology-independent framework for modelling halo properties in simulations of varying resolution. We plan to address this in future work.

In this work we have focused on enhancing the LR halo properties. However, other applications are possible. Fast approaches to numerical simulations \citetext{\citealt{2002MNRAS.329..629S},\citealt{2002MNRAS.331..587M},\citealt{2015MNRAS.450.1856A} and \citealt{2019MNRAS.483L..58B}} can now be improved with \hs to preserve the multi-dimensional AB found in full numerical simulation. This is possible because \hs only requires the large-scale structures and tidal environment as input. This is a very promising application, as internal halo properties are unavailable for the entire dynamic range of the approximate gravity solvers, as has been demonstrated by \citet{2024A&A...692A..32B}. \hs also has the potential to be used to determine the effect that baryons have on dark matter haloes and, possibly, to directly connect galaxies with haloes. The framework is also adaptable for forward modelling approaches to galaxy clustering. This work highlights the potential of ML methods to recover missing small-scale physics in simulations that are limited by resolution.

\begin{acknowledgements}
We thank Aseem Paranjape, Ravi Sheth, Martín de los Ríos, Ángeles Moliné, Sergio Contreras, Francisco-Shu Kitaura, Adrián Gutiérrez, Bernhard Vos Ginés, Santiago Avila and Shadab Alam for the fruitful discussions we have had with them.  
SR and VGP have been supported by the Atracci\'{o}n de Talento Contract no. 2019-T1/TIC-12702 granted by the Comunidad de Madrid in Spain. VGP is also supported by the Atracci\'{o}n de Talento Contract no. 2023-5A/TIC-28943 granted by the Comunidad de Madrid in Spain.
This work has also been supported by Ministerio de Ciencia e Innovaci\'{o}n (MICINN) under the following research grant: PID2021-122603NB-C21 (VGP, GY).

The {\sc unit} simulations have been run  in the MareNostrum
Supercomputer, hosted by the Barcelona Supercomputing Center,
Spain, under the \textsc{prace} project number 2016163937. The analysis in this work has been carried out in the computing cluster at UAM (\textsc{taurus}).
\end{acknowledgements}

\bibliographystyle{aa} 
\bibliography{reference_papers} 
\begin{appendix}
\section{Percentage of unresolved haloes}
\label{sec:motiv}
\begin{figure}
\centering \includegraphics[width=0.9\linewidth]{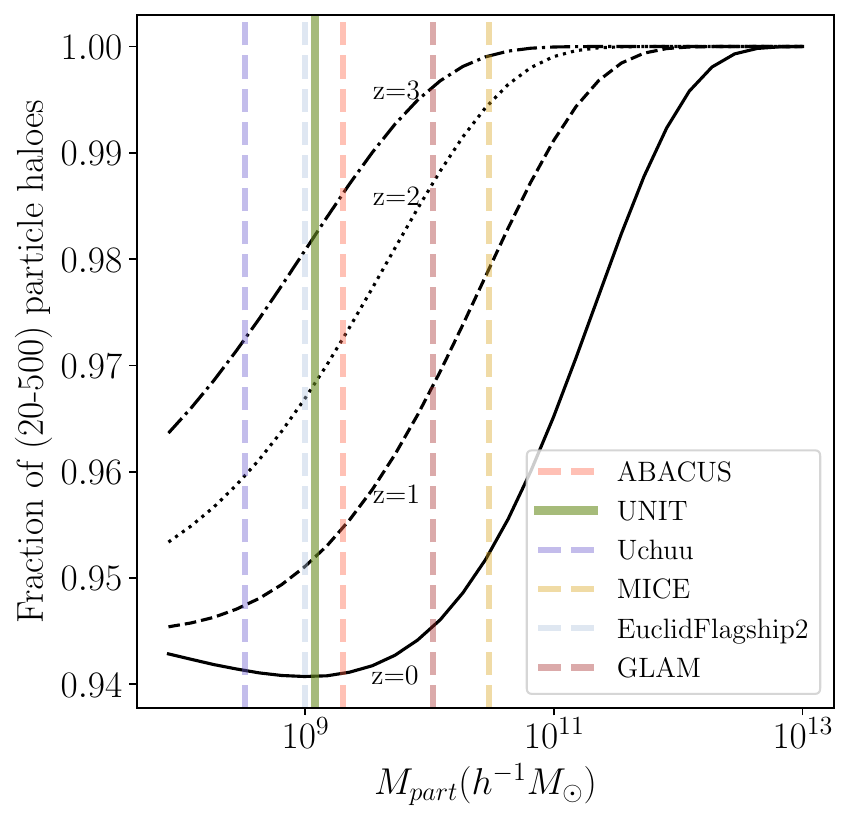}
  \caption{Fraction of the total haloes in a simulation formed by 20 to 500 particles, as a function of the particle resolution, $M_{ \rm part}$, of a simulation. 
  This fraction has been calculated using the \citet{2008ApJ...688..709T} halo mass function with Planck-18 cosmological parameters \citep{2020A&A...641A...6P} and varies for different redshifts (see labels in the figure). The vertical dashed lines of different colours indicate the location of state-of-the-art simulations, as indicated in the legend. The vertical solid green corresponds to {\sc unit}, the simulation used in this work (Sect. \ref{sec:sim}). { Over 90\% of the haloes in these simulations have unresolved halo properties and this number increases for higher redshifts.}}
  \label{fig:fraclowres}
\end{figure}
Over 90\% of dark matter haloes in simulations are unresolved, that is, they do not have converged halo properties (see Fig.~\ref{fig:fraclowres}). The numerical resolution of an N-body simulation sets a limit to the smallest mass of the dark matter halo that can be resolved. Typically, a halo requires a few tens to hundreds of particles for its mass to be well resolved. However, most of the secondary beyond-mass halo properties require at least a few hundred particles (and potentially thousands) to be fully resolved \citep{2021MNRAS.500.3309M}. As the halo mass function decreases rapidly with mass, most of the haloes in a simulation are low mass and thus unresolved. Figure~\ref{fig:fraclowres} shows the fraction of low mass haloes comprising of 20-500 particles as a function of the particle resolution of a simulation. To provide a more realistic view of the problem, we also show in the same figure, state-of-the-art simulations from which galaxy catalogues have been produced. 

\section{Smoothing scale for the environment}
\label{app:justification_for_smoothing_scale}
In this section we provide a self-contained reasoning for the choice of $4\times R_{200b}$ to describe the tidal anisotropy around a halo.\ correlations with smaller smoothing scale peaking for smaller masses and the correlations with larger smoothing scale peaking at higher masses. The black line corresponds to the rank correlation $\rho_{\alpha-\delta}$ at a variable smoothing scale, corresponding to four-times-larger halo radius. Comparing the black line, which forms an envelope around the other coloured lines, it becomes apparent the peak  in rank correlation at any fixed scale is identical to 4 times the radius for that halo mass bin. Since we want to maximally recover the correlations with the large-scale clustering environment, $\alpha_{4 R200b}$ becomes the obvious choice as input parameters in our algorithm as opposed to fixed smoothing scales.
\begin{figure}
    \centering \includegraphics[width=0.9\linewidth]{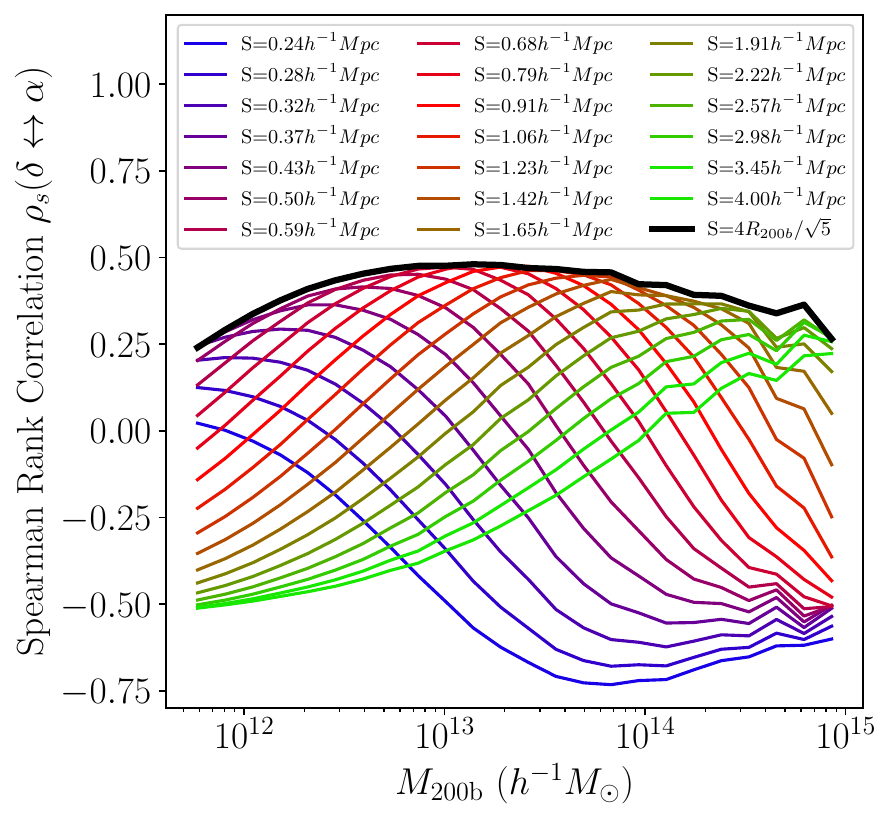}
    \caption{Spearman rank correlation between the tidal anisotropy and the overdensity as a function of halo mass. Different fixed Gaussian smoothing scales, $S$, have been used, as indicated by the legend. The solid black  line indicates the Spearman rank correlation at a variable scale corresponding to the radius of the halo, i.e. $4\cdot R_{200b}/\sqrt{5}$. 
    { Our chosen smoothing scale for the tidal anisotropy, $4\cdot R_{200b}$, corresponds to the maximum correlation for any mass range considered, and the correlation at this scale envelopes the correlations computed at other smoothing scales.} 
    }
    \label{fig:justification_smoothingscale}
\end{figure}

\subsection{Mean and scatter of halo properties}\label{sec:mean}
\begin{figure}
    \centering
    \includegraphics[width=0.8\linewidth]{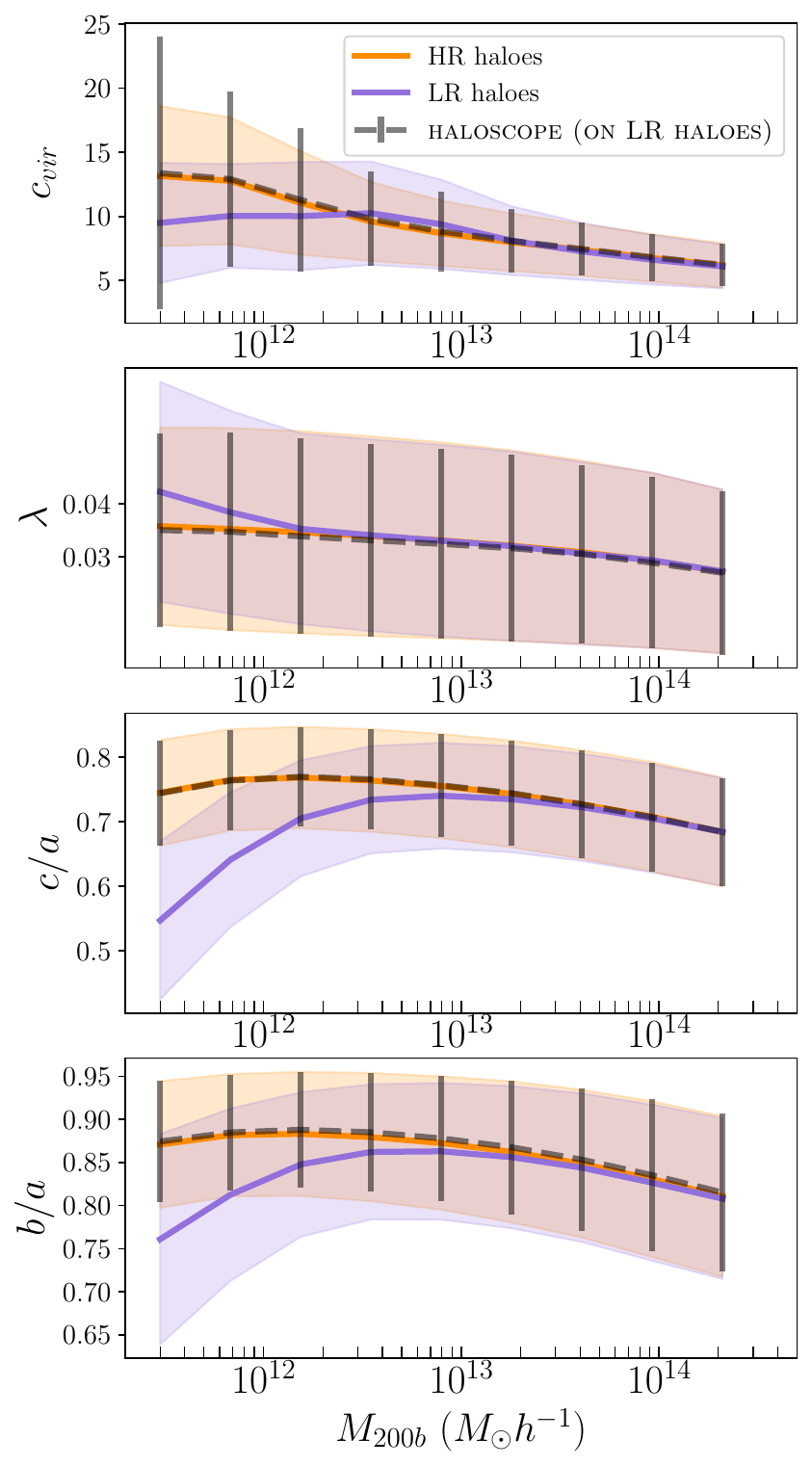}
    \caption{Median halo property (from top to bottom: concentration, spin, and shape) as a function of halo mass for HR (orange), LR (purple), and LR$+$\hs (grey) haloes. The shaded regions show the $68$\ percentile. {Our method recovers the HR median and $1\sigma$ scatter.}}
    \label{fig:prelimresult}
\end{figure}

The numerical convergence problems for LR simulations, can create an offset for the mean and the scatter of properties of low mass haloes. Most methods in the literature aiming to address convergence problems have aimed to reproduce these two statistics, the mean and the scatter. As we can see in Fig.~\ref{fig:prelimresult}, when we apply \hs to the LR haloes we can recover both the mean and the $1\sigma$ scatter measured for the HR haloes.

\section{Halo masses}\label{sec:alternatemass}
\begin{figure}
    \centering
\includegraphics[width=0.85\linewidth]{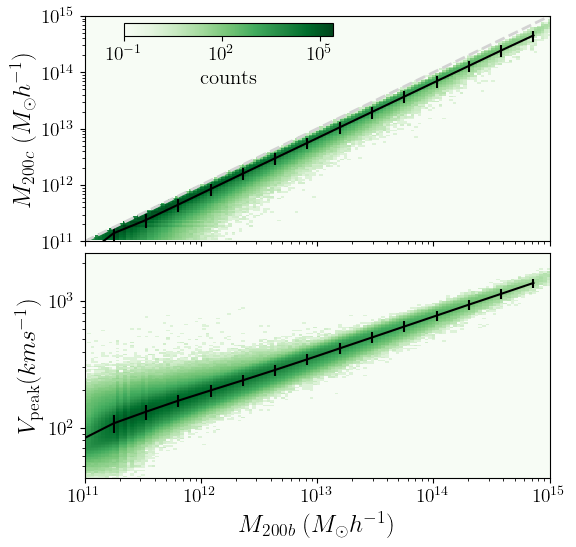}
    \caption{Correlation between $M_{200b}$ and $M_{200c}$ (top panel) and $V_{peak}$ (bottom panel). Black lines show the median and standard deviation. The number of haloes are shown as coloured regions, following the colour bar. The dashed grey line shows the one-to-one relation. { All definitions of the halo mass are highly correlated}, although there is substantial scatter in the case of $V_{peak}$.}
    \label{fig:massdefn}
\end{figure}
In this work we used $M_{\rm 200b}$ as our default definition for the halo mass. $M_{\rm 200b}$ is the mass enclosed inside 200 $\times$ the background density. Another widely used mass definition is $M_{\rm 200c}$, the mass enclosed inside 200 $\times$ the critical density. The peak circular velocity over the accretion history of a halo, $V_{\rm peak}$, has also been widely used in the literature as a proxy for halo masses. 

These three quantities, $M_{\rm 200b}$, $M_{\rm 200c}$, and $V_{\rm peak}$, are correlated (Fig. \ref{fig:massdefn}). The correlation is tighter between $M_{\rm 200b}$ and $M_{\rm 200c}$. In both cases, the scatter increases for smaller values of $M_{\rm 200b}$.

\subsection{Improving the low-mass halo mass function}\label{sec:improve_mass}
\begin{figure}
    \centering
\includegraphics[width=0.85\linewidth]{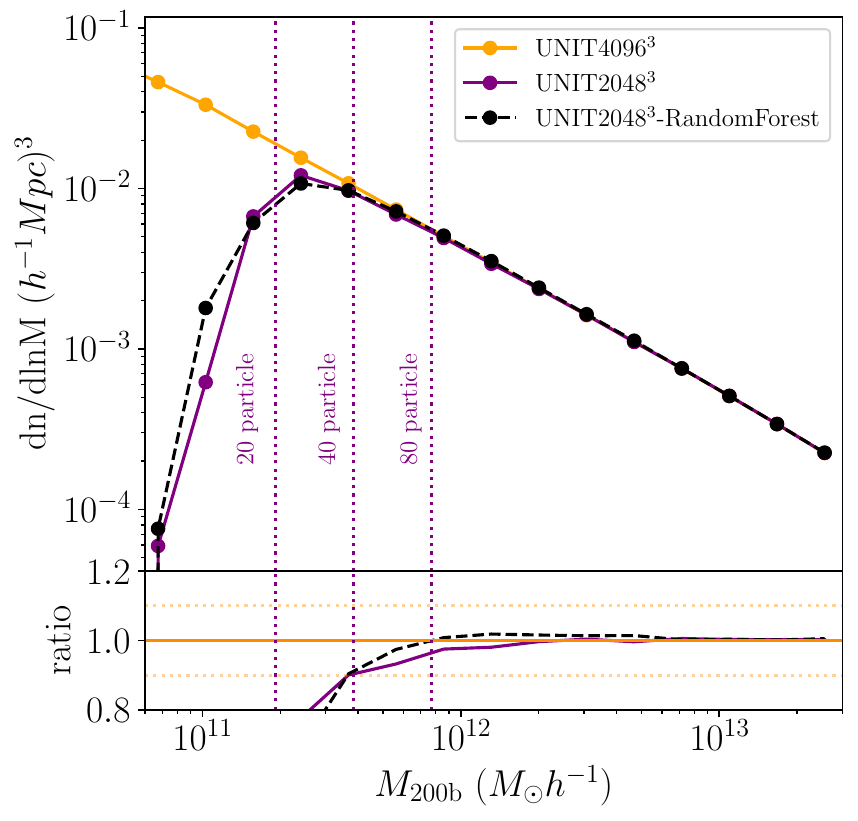}
    \caption{Halo mass function for the LR (solid purple lines) and the HR simulation (solid orange lines). The mass function of the LR simulation is within $10$\% of HR simulation for LR haloes greater than $40$ particles as can be seen from the solid purple line in the bottom panel. The black dashed line corresponds to the LR haloes after correcting their masses with the HR counterparts with a simple 1-1 matching. After the 1-1 matching, and training with RF, the LR halo number density still drops compared to the HR, for haloes with fewer than $40$ particles.  }
    \label{fig:massfunction}
\end{figure}

The primary property of a dark matter halo is its mass. In simulations, there are several definitions and proxies for the mass of the halo due to the lack of consensus on a well-defined halo boundary. The convergence in halo mass is generally much better than for other secondary properties, while the secondary properties require a few hundred particles for convergence, mass convergence can be achieved at a few tens of particles. Yet further improvements in the halo mass function are desirable since model catalogues for large-scale surveys seek to maximise the dynamic range offered by a simulation \citetext{\citealt{2022MNRAS.510L..29A} and \citealt{2014MNRAS.442.3256A}}. Here we improve the mass function by applying the mass correction method in \citet{2022MNRAS.513.4318F}, based on the random forest (RF) technique. 
We followed the hyperparameters for the training algorithm as prescribed in their work, except the number of estimators was set to range between 1 and 10. The other difference is that they use haloes from \textsc{subfind} while we used \textsc{rockstar} haloes.

The LR mass function converges to the HR one, with sub-percent accuracy at high masses (Fig. \ref{fig:massfunction}). At lower masses, the LR halo number density drops compared to the HR. The number of LR haloes with 40 particles is $10$\% below the HR number.

We corrected the masses of LR haloes  using one-to-one matching and an RF algorithm. This correction increases the completeness of the halo catalogue at higher masses (Fig. \ref{fig:massfunction}). However, haloes with fewer than 40 particles are still lost. We find similar results when we perform the 1-1 matching with the alternative mass definitions introduced in Appendix \ref{sec:alternatemass}. We also experimented with alternate methods such as XGboost and  NGboost and conclude that they only bring a marginal improvement over the RF algorithm.

\section{Mass bin ranges}\label{sec:massbins}

\begin{table}[htbp]
\caption{KS statistic comparing 1D distributions of the LR$+$\hs halo properties with respect to the HR ones in different mass bins, as indicated in the first two columns.}
\centering
\footnotesize
\begin{tabular}{|c|c||c|c|c|c|}
\hline
\multicolumn{2}{|c||}{$M_{200b}({\rm M}_{\odot} h^{-1})$}  & {$c_{vir}$} & {$\lambda$} & { $c/a$ } & {$b/a$} \\
Min. & Max.  &  &  &  &  \\
\hline
2 $\times 10^{11}$ & 4.5 $\times 10^{11}$ & 0.012 & 0.012 & 0.015 & 0.023 \\
\hline
4.5 $\times 10^{11}$ & 1$\times 10^{12}$ & 0.012 & 0.012 & 0.014 & 0.025 \\
\hline
1 $\times 10^{12}$&  $2.3 \times 10^{12}$ & 0.012 & 0.012 & 0.014 & 0.026 \\
\hline
$2.3 \times 10^{12}$ & $5.3 \times 10^{12}$ & 0.01 & 0.011 & 0.013 & 0.027 \\
\hline
$5.3 \times 10^{12}$ & $1.2 \times 10^{13}$ & 0.011 & 0.01 & 0.012 & 0.027 \\
\hline
$1.2 \times 10^{13}$ & $2.7 \times 10^{13}$ & 0.01 & 0.01 & 0.011 & 0.022 \\
\hline
2.7 $\times 10^{13}$ & $6.15 \times 10^{13}$ & 0.009 & 0.009 & 0.01 & 0.019 \\
\hline
$6.15 \times 10^{13}$ & $1.4 \times 10^{14}$ & 0.008 & 0.008 & 0.008 & 0.015 \\
\hline
$1.4 \times 10^{14}$ & $3.2 \times 10^{14}$ & 0.008 & 0.007 & 0.007 & 0.005 \\
\hline
\end{tabular}

\tablefoot{All the reported values are below the null hypothesis threshold. There is a good agreement between the distributions of LR$+$\hs halo properties and those from the HR simulation} 
\label{tab:kstest}
\end{table}

\begin{figure*}
\includegraphics[width=18cm]{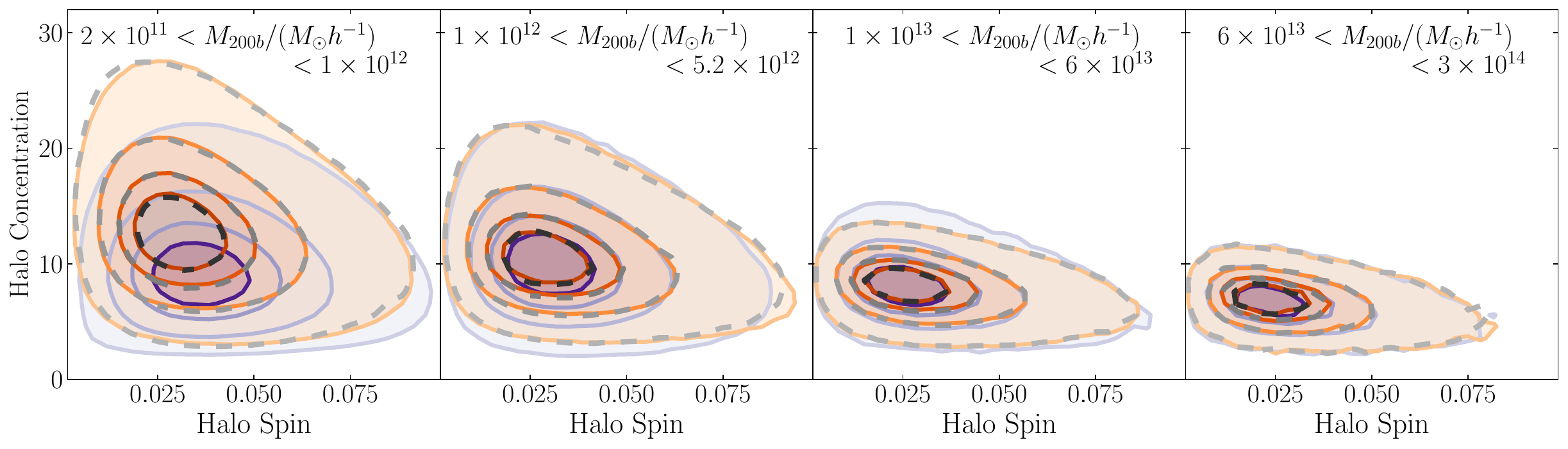}
\caption{Same as the top panel of Fig.~\ref{fig:bivariate-distribution} but for different mass bins. The right panel corresponds to massive haloes, with at least 6000 particles. The orange contours are for the HR haloes and the purple ones for the LR haloes. Dashed grey contours show the result of applying \hs to the LR haloes. {The differences between HR and LR decrease with halo mass, as halo properties can be resolved when haloes contain a large enough number of particles}.
}
\label{fig:bivariate-distribution-massbins}
\label{fig:massbin}
\end{figure*}

\hs is applied in different mass bins (two first columns in \autoref{tab:kstest}). This implies that $\vec{\bar{\Sigma}}$ and $\vec{\bar{\mu}}$ in Eqs.~\ref{eq:mubar} and \ref{eq:sigmabar} are recomputed for different mass ranges.

The differences between LR bivariate distributions and those for the HR haloes, decrease with increasing halo mass. This is expected, as halo properties start to be resolved in the LR simulation above a certain mass limit, for which enough haloes contain enough particles. This is shown in Fig. \ref{fig:bivariate-distribution-massbins} for the case of halo concentration versus halo spin. However, similar results are obtained for the other halo properties under study.

To quantify the performance of \hs recovering the HR halo property distributions, we used the Kolmogorov–Smirnov (KS) test from the \textsc{scipy} module for the distribution of individual halo properties. In this case, the null hypothesis corresponds to the two samples belonging to the same distribution. 

We computed the KS statistic in different mass bins (\autoref{tab:kstest}). We limited our sample size to $10000$ haloes in each mass range to prevent heightened sensitivity in our statistical tests for large sample sizes and to prevent frequent null hypothesis rejection \citep{41f077eb-868d-3493-9b59-c0228124bc2a}. The $10000$ haloes have been sampled randomly in each mass bin under study. Since p-value indicates the strength of evidence against the null hypothesis, we chose a p-value $\leq$ 0.001 to reject the null hypothesis, which corresponds to a KS statistic $\geq$ 0.0274. 

All the studied cases are consistent with the null hypothesis (\autoref{tab:kstest}), and thus, we can consider both the HR and LR$+$\hs halo properties to have the same distributions.

\end{appendix}
\end{document}